# Optimum Power Allocations for Fading Decode-and-Forward Relay Channel

Arash Gholami Davoodi, Mohammad Javad Emadi, Mohammad Reza Aref

*Abstract*—**In this paper optimum power allocations are derived for a fading decode-and-forward full-duplex relay channel, analytically. Individual power constraints for the source and the relay are assumed and the related optimization problem is analyzed for two scenarios. First, optimization is taken over the source power, the relay power, and the correlation coefficient between the transmitted signals of the source and the relay. Then, for a fixed value of correlation coefficient, the optimization problem is analyzed. It is also proved that the optimization problem is convex for these scenarios. The problem is analyzed in the three possible cases, and optimum power allocations are derived in closed-form for each case. Finally, theoretical results are evaluated through simulations, and for each scenario; we also show how the transition between the three cases takes place.**

## I. INTRODUCTION

Improving reliability and throughput of wireless networks have been one of the important challenges, in recent decades. Utilizing relay nodes in networks would be beneficial and has become a significant area of research. The relay channel has been introduced in [1]-[3] and comprehensively analyzed in [4]-[5].

The authors are with the ISSL Lab, Electrical Engineering Department, Sharif University of Technology, Tehran, Iran. {Arash.G.Davoodi@gmail.com, corresponding author: emadi@ee.sharif.edu, and aref@sharif.edu). This work was supported in part by the Iranian National Science Foundation (INSF) under Contract No. 88114.46-2010 and by the Iran Telecom Research Center (ITRC) under Contract No. 500.18495.

Moreover, three achievable rates have been established for this channel utilizing the well-known relaying strategies: amplify-and-forward (AF), compress-and-forward (CF), and decode-and-forward (DF).

Another important issue in wireless networks is the problem of determining optimum power allocations for the users. Many power allocation algorithms are based on convex optimization techniques, because of the availability of powerful analytic and numerical tools [6]. Convex optimization approaches have been considered for a wide variety of network information theory problems; a review of the related literature is given in [7]-[11].

Resource allocation for relay channels and some other network information theory have been studied in [12]-[15]. In all of these studies, it is assumed that the source and relay nodes are subject to total power constraint. In [16], the OFDM multiple access relay channel has been considered and maximization of the sum-rate under sum-power constraint on the power allocation of the different carriers for the users and the relay has been analyzed. In [17], efficient power allocation schemes for multi-user wireless AF relay systems has been developed for some scenarios. Optimum power allocations of the two-way AF and DF relay channel under fairness constraint has been studied in [18] and the maximum sum-rate has been derived.

On the other hand, in [19], a number of power allocations for the relay channel under individual power constraints of the source and relay has been analyzed where the assumptions are more practical in view of wireless networks; also, parallel relay channel has been studied and the cut-set upper bound derived. Moreover, a lower bound on the capacity has been established using the partial DF strategy and the optimum power allocations for the fading parallel relay channel has been characterized for the asynchronous mode. Furthermore, for both the full-duplex and half-duplex cases, power allocations have been analyzed, and for the synchronous case the problem has been partially investigated.

In [20], the full-duplex and half-duplex Gaussian relay channels with correlated noises at the relay and receiver has been considered and using the AF, DF, and CF strategies, inner bounds and an outer bound for the capacity have been established. It has been shown that the achievable rates employing the AF and CF strategies depend on the correlation coefficient of the noises and these approaches can potentially exploit such extra information. But the achievable rate of the DF strategy is independent of the correlation coefficient of the noises. In addition, for fixed channel gains in every node and under sum-power constraint of the source and relay, optimum power allocations have been derived for the achievable rates employing the CF and AF strategies. It is worthy to note that "none of the three relay schemes always outperforms the other two for all values of the correlation coefficients of the noises".

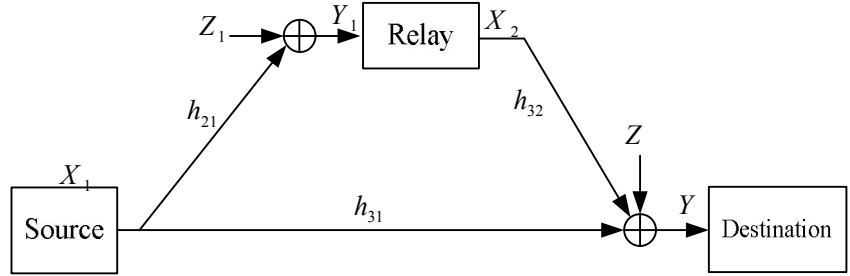

Fig. 1. The fading Gaussian relay channel [20].

Authors in [19] conjecture that deriving optimum power allocations for the fading full-duplex Gaussian relay channel under the DF strategy would be too complex, and one needs to do a brute-force search to find the set of optimum power allocations for this problem. The main result of this paper is to derive the optimum power allocations in closed-form, analytically. First, we obtain optimum power allocations for a general case: optimum value of $\rho_x$ (the correlation coefficient between the transmitted signals of the source and the relay), and the optimum power allocations for the source and the relay are derived in closed-form. Then, some implications of the results for the structure of power allocation problem in Rayleigh fading Gaussian relay channels are derived and discussed. Then, to reduce the complexity of the transmitter, we assume a fixed value of $\rho_x$ and then derive the optimum power allocations in closed-form. Finally, other implications of the results are discussed.

The remainder of the paper is organized as follows: The system model and problem statements are introduced in Section II. Optimum power allocations are established in Section III. Numerical results and implications are presented in Section IV, and finally we conclude the paper in Section V.

## II. SYSTEM MODEL AND PROBLEM STATEMENT

### A. System Model

The full-duplex mode of the fading Gaussian relay channel was considered in [20] wherein the relay can transmit and receive signals at the same time over the same frequency band. In full-duplex mode, the channel is described by the following equations:

$$\begin{aligned} Y_1(i) &= h_{21}X_1(i) + Z_1(i) \\ Y(i) &= h_{31}X_1(i) + h_{32}X_2(i) + Z(i), \end{aligned} \quad (1)$$

where, $i$ indicates the $i$th time slot, $X_1(i)$ and $X_2(i)$ are the channel input signals from the source and the relay with power constraints $\overline{P_1}$ and $\overline{P_1}$, respectively, and are arbitrarily correlated with correlation coefficient $\rho_x$. $Y_1(i)$ and $Y(i)$ are the received signals at the relay and the destination, respectively. $Z_1(i)$ and $Z(i)$ are arbitrarily correlated Gaussian noises with correlation coefficient $\rho_z$ and variances of $N_1$ and $N_1 + N_2$, respectively. $h_{21}$, $h_{31}$ and $h_{32}$ are channel gains between the nodes, see Fig. 1. In the following, it is also assumed that the channel gains are known at all the nodes and the variation of the channel gains are slow enough to track the fading parameters precisely at the relay and at the transmitter and yet fast enough to ensure that the long-term ergodic properties of the channel are observed within the transmission blocks. Individual power constrains for the source and the relay are considered as follows:

$$\frac{1}{n}\sum_{i=1}^{n} x_{1i}^2(w) \leq \overline{P_1}, \qquad w \in \{1,2,\dots,M\}, \quad (2)$$

$$\frac{1}{n}\sum_{i=1}^{n} x_{2i}^2(y_{11}, y_{12}, \dots, y_{1i-1}) \leq \overline{P_2}, \qquad \mathbf{y}_1^n \in \mathbb{R}^n, \quad (3)$$

where $w$ is the transmitted message of the source.

### B. Problem Statements

An achievable rate of the channel using the DF strategy was established in [20]. In this paper, we analyze two scenarios for deriving optimum power allocations. First, optimization is taken over the source and relay powers, i.e., $P_1$ and $P_2$, and the correlation coefficient $\rho_x$. Then, for a fixed $\rho_x$, optimization is only taken over the source and relay powers.

Employing the achievable rate of [20], the following two definitions are introduced.

***Definition 1* [Optimization over power allocations and $\rho_x$]:** Using achievable rate derived in [20] and suitable change of variables as

$$P_r := (1 - \rho_x^2)P_1, \quad P_s := \rho_x^2 P_1, \tag{4}$$

the optimization problem is rewritten as

$$R_{\text{full-duplex,DF}} = \max_{P_r, P_s, P_2} \min\{E\{R_1(P_r, P_s, P_2, h_{31}, h_{32})\}, E\{R_2(P_r, h_{21})\}\}, \tag{5}$$

where expectation is taken over the channel gains, and $R_1(.)$ and $R_2(.)$ are

$$R_1(P_r, P_s, P_2, h_{31}, h_{32}) = C\left(\frac{(P_r + P_s)h_{31}^2 + P_2 h_{32}^2 + 2\sqrt{P_s P_2} h_{31} h_{32}}{N_1 + N_2}\right) \tag{6}$$

$$R_2(P_r, h_{21}) = C\left(\frac{P_r h_{21}^2}{N_1}\right), \tag{7}$$

where $C(x) := \frac{1}{2}\log(1 + x)$.

***Definition 2* [Optimization over power allocations for fixed $\rho_x$]:** For a fixed $\rho_x$, the optimization problem (5) is rewritten as

$$R_{\text{full-duplex,DF}} = \max_{P_1, P_2} \min\{E\{R_1(P_1, P_2, \rho_x, h_{31}, h_{32})\}, E\{R_2(P_1, \rho_x, h_{21})\}\}, \tag{8}$$

and $R_1(.)$ and $R_2(.)$ are

$$R_1(P_1, P_2, \rho_x, h_{31}, h_{32}) = C\left(\frac{P_1 h_{31}^2 + P_2 h_{32}^2 + 2\rho_x \sqrt{P_1 P_2} h_{31} h_{32}}{N_1 + N_2}\right), \tag{9}$$

$$R_2(P_1, \rho_x, h_{21}) = C\left(\frac{(1 - \rho_x^2)P_1 h_{21}^2}{N_1}\right). \tag{10}$$

In this paper, we prove that the optimization problems (5) and (8) are both convex, and analytically derive the optimum power allocations in closed-form for the two scenarios.

III. MAIN RESULTS

Before we present the optimum power allocation results for the two scenarios, in the following subsection, some properties of the optimization problems are discussed in lemmas 1-3.

*A. Some Properties of the Optimization Problems*

Since the optimization problems (5) and (8) contain expectation operators, in the following lemma we prove that, if certain conditions are met, it suffices to solve the optimization problem regardless of the expectation operator.

*Lemma 1:* Consider a general optimization problem as:

$$\max_{p_k(H), 1 \leq k \leq K} \mathrm{E}_H\{F(p_1(H), \ldots, p_K(H), H)\} \tag{11}$$

$$\text{s.t.} \quad \mathrm{E}_H\{G_i(p_1(H), \ldots, p_K(H), H)\} \leq \bar{G}_i, 1 \leq i \leq I, \tag{12}$$

where $H \coloneqq \{h_j, 1 \leq j \leq J\}$ is a set of random variables and the expectation is taken over the set; $I, J$, and $K$ are arbitrary positive integer numbers; all $p_i(H)$ are arbitrary positive functions; $\bar{G}_i$'s are arbitrary positive numbers which can be denoted as the average power allocation constraint, $F(p_1(H), \ldots, p_K(H), H)$ and all $G_i(p_1(H), \ldots, p_K(H), H)$ are convex functions with respect to $p_k(H)$. If the Lagrangian function: $F(p_1(H), \ldots, p_K(H), H) - \sum_{1 \leq i \leq I} \lambda_i G_i(p_1(H), \ldots, p_K(H), H)$ has optimum solutions $p_k^*$, then all $p_k^*$ are also solutions of the original optimization problem (11) and $\lambda_i$'s are computed from the constraints (12).

*Proof*: It is easy to verify that Slater's condition is satisfied [6]. Hence, there exist constants $\lambda_i$, and the solution of (11) is the same as the solution of the following optimization problem:

$$\max_{p_k(H), 1 \leq k \leq K} \left\{ \mathrm{E}_H\{F(p_1(H), \ldots, p_K(H), H)\} + \sum_{1 \leq i \leq I} \lambda_i (\bar{G}_i - \mathrm{E}_H\{G_i(p_1(H), \ldots, p_K(H), H)\}) \right\}. \tag{13}$$

As the functions $F(p_1(H), \ldots, p_K(H), H)$ and $G_i(p_1(H), \ldots, p_K(H), H)$ are convex functions with respect to $p_k(H)$, the optimization problem is also convex. Thus, the Kuhn-Tucker conditions are sufficient to establish the optimum solution, and the optimum $\lambda_i$ is unique and derived from the constraints (12).

Hence, the Lagrangian function is given by

$$\mathcal{L}(p_1(H), \ldots, p_K(H), H) := F(p_1(H), \ldots, p_K(H), H) - \sum_{1 \leq i \leq I} \lambda_i (G_i(p_1(H), \ldots, p_K(H), H) - \bar{G}_i). \quad (14)$$

If $p_k^*(H)$, $k = 1, \ldots, K$ are optimum solutions of $\mathcal{L}(.)$ for each state $H$, then,

$$\mathcal{L}(p_1^*(H), \ldots, p_K^*(H), H) \geq \mathcal{L}(p_1(H), \ldots, p_K(H), H), \text{ for fixed } \lambda_i \ \forall p_k(H), H. \quad (15)$$

Taking the expected value of both sides of (15) and applying the linearity of expectation, we have

$$E_H\{\mathcal{L}(p_1^*(H), \ldots, p_K^*(H), H)\} \geq E_H\{\mathcal{L}(p_1(H), \ldots, p_K(H), H)\}, \ \forall p_k(H), H. \quad (16)$$

Thus, the functions $p_k^*(H)$ are also the optimum solutions of the original optimization problem (11). ∎

In order to utilize Lemma 1 throughout the paper, in the following lemma, we prove that the optimization problems (5) and (8) are convex.

***Lemma 2:*** The two optimization problems (5) and (8) are convex in the set of all possible positive power allocations.

***Proof:*** Since the convexity of power constraints are obvious, only the convexity of the objective functions in equations (6), (7), (9) and (10) must be investigated. First, to prove the convexity of $R_1(P_r, P_s, P_2, h_{31}, h_{32})$, for $0 \leq \alpha \leq 1$ and two given points $(P_{r1}, P_{s1}, P_{21})$ and $(P_{r2}, P_{s2}, P_{22})$, writing the convexity definition, (17) is concluded.

$$\alpha C\left(\frac{(P_{r1}+P_{s1})h_{31}^2 + P_{21}h_{32}^2 + 2\sqrt{P_{s1}P_{21}}h_{31}h_{32}}{N_1+N_2}\right) + (1-\alpha)C\left(\frac{(P_{r2}+P_{s2})h_{31}^2 + P_{22}h_{32}^2 + 2\sqrt{P_{s2}P_{22}}h_{31}h_{32}}{N_1+N_2}\right)$$
$$\overset{(a)}{\leq} C\left(\frac{\big(\alpha(P_{r1}+P_{s1}) + (1-\alpha)(P_{r2}+P_{s2})\big)h_{31}^2 + (\alpha P_{21} + (1-\alpha)P_{22})h_{32}^2 + 2\big(\alpha\sqrt{P_{s1}P_{21}} + (1-\alpha)\sqrt{P_{s2}P_{22}}\big)h_{31}h_{32}}{N_1+N_2}\right),$$
$$(17)$$

where (a) follows from the convexity of the function $C(.)$. On the other hand, it is clear that

$$\sqrt{P_{s1}P_{21}} + \sqrt{P_{s2}P_{22}} \leq \sqrt{(P_{s1}+P_{s2})(P_{21}+P_{22})} \quad (18)$$

Therefore, using inequality (18) in (17), then (19) is obtained.

$$\alpha C\left(\frac{(P_{r1}+P_{s1})h_{31}^2 + P_{21}h_{32}^2 + 2\sqrt{P_{s1}P_{21}}h_{31}h_{32}}{N_1+N_2}\right) + (1-\alpha)C\left(\frac{(P_{r2}+P_{s2})h_{31}^2 + P_{22}h_{32}^2 + 2\sqrt{P_{s2}P_{22}}h_{31}h_{32}}{N_1+N_2}\right)$$

$$\leq C\left(\frac{(\alpha(P_{r1}+P_{s1})+(1-\alpha)(P_{r2}+P_{s2}))h_{31}^2 + (\alpha P_{21}+(1-\alpha)P_{22})h_{32}^2 + 2\sqrt{(\alpha P_{s1}+(1-\alpha)P_{s2})(\alpha P_{21}+(1-\alpha)P_{22})}h_{31}h_{32}}{N_1+N_2}\right)$$

(19)

Thus, $R_1(P_r, P_s, P_2, h_{31}, h_{32})$ is a convex function. Similarly, it is proved that $R_2(P_r, h_{21})$, $R_1(P_1, P_2, \rho_x, h_{31}, h_{32})$ and $R_2(P_1, \rho_x, h_{21})$ are also convex functions for all possible positive power allocations. It is worth noting that, it is possible to prove the convexity of the functions by proving that the relative Hessian matrices are semi-negative definite matrices [6]. Finally, as the minimum of two convex functions is also a convex function, the two optimization problems (5) and (8) are convex in the set of power allocations. ∎

We use the following definitions hereinafter.

***Definition 3***: Surface $J$ is defined as a confluence of two surfaces $\mathrm{E}\{R_1(.)\}$ and $\mathrm{E}\{R_2(.)\}$.

***Definition 4***: $P_k^{c_3}$ for $k, i = 1,2$ are defined as the power allocations which maximize the surface $J$.

***Definition 5***: $P_k^{c_i}$ for $k, i = 1,2$ are defined as the power allocations which maximize the function $\mathrm{E}\{R_i(.)\}$.

***Definition 6***: $R_{i_{P_1=P_1^{c_i}, P_2=P_2^{c_i}}}$ for $k, i = 1,2$ are defined as the rate $R_i$ in which the power allocations $P_k$ are substituted by the terms $P_k^{c_i}$ for $k = 1,2$.

In the following lemma, we prove that solving the optimization problems (5) and (8) fall into three cases.

***Lemma 3:*** To analyze the optimization problem (5), three cases can be considered.

**Case 1.** If

$$\mathrm{E}\{R_2(P_r, P_s, h_{21})\}_{P_r=P_r^{c_1}, P_s=P_s^{c_1}} \geq \mathrm{E}\{R_1(P_r, P_s, P_2, h_{31}, h_{32})\}_{P_r=P_r^{c_1}, P_s=P_s^{c_1}, P_2=P_2^{c_1}},$$

then

$$\max_{P_r, P_s, P_2} \min(\mathrm{E}\{R_1(.)\}, \mathrm{E}\{R_2(.)\}) = \mathrm{E}\{R_1(.)\}_{P_r=P_r^{c_1}, P_s=P_s^{c_1}, P_2=P_2^{c_1}},$$

and $P_r^{c_1}, P_s^{c_1}$ and $P_2^{c_1}$ are the optimum power allocations. So, the Lagrangian function for deriving the

optimum power allocation is given by

$$\mathcal{L} = \mathrm{E}\{R_1(.)\} + \lambda_1(\overline{P_1} - \mathrm{E}\{P_r + P_s\}) + \lambda_2(\overline{P_2} - \mathrm{E}\{P_2\}). \tag{20}$$

This optimization problem has a unique global maximum which is obtained from the Kuhn-Tucker conditions, and these conditions are sufficient to derive the maximum. $\lambda_k, k = 1,2$ are uniquely derived from the two power constraints.

**Case 2.** If

$$\mathrm{E}\{R_2(.)\}_{P_r=P_r^{c_2},P_s=P_s^{c_2}} \leq \max_{P_2} \mathrm{E}\{R_1(.)\}_{P_r=P_r^{c_2},P_s=P_s^{c_2}},$$

then

$$\max_{P_r,P_s,P_2} \min(\mathrm{E}\{R_1(.)\}, \mathrm{E}\{R_2(.)\}) = \mathrm{E}\{R_2(.)\}_{P_r=P_r^{c_2},P_s=P_s^{c_2}}$$

and $P_r^{c_2}, P_s^{c_2}$ and $P_2^{c_2}$ are the optimum power allocations. In this case, the Lagrangian function for deriving the optimum power allocation is given by

$$\mathcal{L} = \mathrm{E}\{R_2(.)\} + \lambda_1(\overline{P_1} - \mathrm{E}\{P_r + P_s\}). \tag{21}$$

This optimization problem has a unique global maximum which is obtained from the Kuhn-Tucker conditions, and these conditions are sufficient to derive the maximum. $\lambda_1$ is uniquely derived from the power constraint.

**Case 3.** If none of the Cases 1 and 2 ocurrs, then $\max_{P_r,P_s,P_2} \min(\mathrm{E}\{R_1(.)\}, \mathrm{E}\{R_2(.)\})$ occurs on the surface $J$ where the two surfaces $\mathrm{E}\{R_1(.)\}$ and $\mathrm{E}\{R_2(.)\}$ are equal and maximum and $P_r^{c_3}, P_s^{c_3}$ and $P_2^{c_3}$ are the optimum power allocations. Therefore, the Lagrangian function for deriving the optimum power allocation in this case is written as

$$\mathcal{L} = \mathrm{E}\{R_1(.)\} + \lambda_3(\mathrm{E}\{R_2(.)\} - \mathrm{E}\{R_1(.)\}) + \lambda_1(\overline{P_1} - \mathrm{E}\{P_r + P_s\}) + \lambda_2(\overline{P_2} - \mathrm{E}\{P_2\}), \tag{22}$$

where $0 \leq \lambda_3 \leq 1$ and this optimization has unique global maximum which is obtained from the Kuhn-Tucker conditions, and these conditions are sufficient to derive it. $\lambda_j, j = 1,2,3$ are uniquely derived from the power constraints and the equality $\mathrm{E}\{R_2(.)\} = \mathrm{E}\{R_1(.)\}$.

*Proof:* **See Appendix.**

Note that, using Lemma 3 and replacing $P_r$ and $P_s$ by $(1-\rho_x^2)P_1$ and $\rho_x^2 P_1$, respectively, three cases occur for solving the optimization problem (8).

## B. Optimum Power Allocations

In this subsection, optimum power allocations for the two scenarios are established in the theorems 1 and 2.

### 1) Optimum Power Allocations and Optimum $\rho_x$

To derive the optimum power allocations of the fading Gaussian relay channel in a general form, the optimization problem (5) is solved and the results are given in Theorem 1.

***Theorem 1:*** The optimum power allocations of the fading Gaussian Relay, i.e., $P_r(.)$, $P_s(.)$ and $P_2(.)$ are derived for the three cases as follows:

**Case 1.** If

$$E\{R_2(P_r, P_s)\}_{P_r=P_r^{c_1}, P_s=P_s^{c_1}} \geq E\{R_1(P_r, P_s, P_2)\}_{P_r=P_r^{c_1}, P_s=P_s^{c_1}, P_2=P_2^{c_1}},$$

the optimum power allocations are

$$P_r^{c_1} = 0$$

$$P_s^{c_1}(h_{31}, h_{32}, \Gamma) = \left( \frac{\frac{h_{31}^2 + \Gamma h_{31} h_{32}}{\lambda_1} - N_1 - N_2}{h_{31}^2 + \Gamma^2 h_{32}^2 + 2\Gamma h_{31} h_{32}} \right)^+ \quad (23)$$

$$P_2^{c_1}(h_{31}, h_{32}, \Gamma) = \Gamma^2 P_s^{c_1}(h_{31}, h_{32}, \Gamma)$$

$$\Gamma = \frac{\lambda_1 h_{32}}{\lambda_2 h_{31}},$$

where

$$(x)^+ = \begin{cases} x, & \text{If } x \geq 0 \\ 0, & \text{If } x < 0 \end{cases}.$$

It is clear that, if $P_r^{c_1}=0$ is substituted in (7), then $E\{R_2(.)\}_{P_r=0, P_s=P_s^{c_1}} = 0$ and $E\{R_1(.)\}_{P_r=P_r^{c_1}, P_s=P_s^{c_1}, P_2=P_2^{c_1}} \leq 0$. Consequently, this case does not happen.

**Case 2.** If

$$E\{R_2(.)\}_{P_r=P_r^{c_2},P_s=P_s^{c_2}} \le \max_{P_2} E\{R_1(.)\}_{P_r=P_r^{c_2},P_s=P_s^{c_2}},$$

the optimum power allocations are

$$P_s^{c_2} = 0$$

$$P_r^{c_2}(h_{21}) = \left(\frac{1}{2\lambda_1} - \frac{N_1}{h_{21}^2}\right)^+ \quad (24)$$

$$P_2^{c_2}(h_{21}, h_{31}, h_{32}) = \left(\frac{1}{\lambda_1} - \frac{N_1 + N_2}{h_{32}^2} - \frac{P_r^{c_2} h_{31}^2}{h_{32}^2}\right)^+.$$

**Case 3.** If none of the Cases 1 and 2 occurs, then optimum power allocations of $\max_{P_r,P_s,P_2} \min(E\{R_1(.)\}, E\{R_2(.)\})$ are

$$P_r^{c_3}(h_{21}, h_{31}, h_{32}) = \left(\frac{\lambda_3(h_{31}^2 + \Gamma h_{31} h_{32})}{\lambda_1 \Gamma h_{31} h_{32}} - \frac{N_1}{h_{21}^2}\right)^+$$

$$P_s^{c_3}(h_{21}, h_{31}, h_{32}) = \frac{\left(\frac{(1-\lambda_3)(h_{31}^2 + \Gamma h_{31} h_{32})}{\lambda_1} - N_1 - N_2 - P_r^{c_3} h_{31}^2\right)^+}{h_{31}^2 + \Gamma^2 h_{32}^2 + 2\Gamma h_{31} h_{32}} \quad (25)$$

$$P_2^{c_3}(h_{21}, h_{31}, h_{32}) = \Gamma^2 P_s^{c_3}(h_{21}, h_{31}, h_{32})$$

$$\Gamma = \frac{\lambda_1 h_{32}}{\lambda_2 h_{31}}.$$

***Proof*: See Appendix.**

2) *Optimum Power Allocations for Fixed $\rho_x$*

To reduce the complexity of the transmitter, we assume a fixed $\rho_x$ throughout the transmission block and solve the optimization problem (8).

***Theorem 2*:** The optimum power allocations of the fading Gaussian relay channel, i.e., $P_k(.) \ \forall k = 1,2$, are derived as follows:

**Case 1.** If

$$E\{R_2(P_1, \rho_x, h_{21})\}_{P_1=P_1^{c_1}} \ge E\{R_1(P_1, P_2, \rho_x, h_{31}, h_{32})\}_{P_1=P_1^{c_1}, P_2=P_2^{c_1}},$$

the optimum power allocations are

$$P_1^{c_1}(h_{31}, h_{32}, \Gamma) = \left( \frac{\frac{h_{31}^2 + \rho_x \Gamma h_{31} h_{32}}{\lambda_1} - N_1 - N_2}{h_{31}^2 + \Gamma^2 h_{32}^2 + 2\rho_x \Gamma h_{31} h_{32}} \right)^+ \tag{26}$$

$$P_2^{c_1}(h_{31}, h_{32}, \Gamma) = \Gamma^2 P_1^{c_1}(h_{31}, h_{32}, \Gamma),$$

where $\Gamma$ is defined in Appendix.

**Case 2.** If

$$E\{R_2(.)\}_{P_1=P_1^{c_2}} \le \max_{P_2} E\{R_1(.)\}_{P_1=P_1^{c_2}},$$

the optimum power allocations are

$$P_1^{c_2}(h_{21}) = \left( \frac{1}{2\lambda_1} - \frac{N_1}{(1-\rho_x^2)h_{21}^2} \right)^+$$

$$P_2^{c_2}(h_{21}, h_{31}, h_{32})$$
$$= \frac{1}{h_{32}^2} \left( \left( \frac{-F}{3E} - \frac{1}{3E} \sqrt[3]{\frac{1}{2}\left(A + \sqrt{(A^2 - 4(F^2 - 3EG)^3)^+}\right)} \right. \right.$$
$$\left. \left. - \frac{1}{3E} \sqrt[3]{\frac{1}{2}\left(A - \sqrt{(A^2 - 4(F^2 - 3EG)^3)^+}\right)} \right)^+ \right)^2, \tag{27}$$

where $A$, $E$, $F$ and $G$ are defined in Appendix.

**Case 3.** If none of the Cases 1 and 2 occurs, then the optimum power allocations of $\max_{P_1, P_2} \min(E\{R_1\}, E\{R_2\})$ occur on the surface $J$ and are given by

$$P_1^{c_3}(\lambda_1, \lambda_2, \lambda_3, h_{21}, h_{31}, h_{32}, \Gamma) = \left( \frac{\lambda}{\lambda_1 - \frac{\lambda_2(h_{31}^2 + \rho_x \Gamma h_{31} h_{32})}{h_{32}^2 + \frac{\rho_x}{\Gamma} h_{31} h_{32}}} - \frac{N_1}{(1-\rho_x^2)h_{21}^2} \right)^+ \tag{28}$$

$$P_2^{c_3}(\lambda_1, \lambda_2, \lambda_3, h_{21}, h_{31}, h_{32}, \Gamma) = \Gamma^2 P_1^{c_3}(\lambda_1, \lambda_2, \lambda_3, h_{21}, h_{31}, h_{32}, \Gamma),$$

where $\Gamma$ is defined in Appendix.

***Proof*: See Appendix.**

**Remark:** It is clear that in all the cases of Theorems 1 and 2 the values of $\Gamma$ are always non-negative, except in Case 3 of Theorem 2 that is not clear. As a result, If $\Gamma < 0$ is obtained in that case, it means that

$P_2 = 0$. Therefore, setting $P_2 = 0$ in (8) and, we obtain

$$P_2^{c_3}(.) = 0$$

$$P_1^{c_3}(h_{21}, h_{31}) = \frac{1 - \lambda_1\left(\frac{N_1 + N_2}{h_{31}^2} + \frac{N_1}{h_{21}^2}\right) + \sqrt{\Delta}}{2\lambda_1}, \quad (29)$$

$$\Delta = \left(1 - \lambda_1\left(\frac{N_1 + N_2}{h_{31}^2} + \frac{N_1}{h_{21}^2}\right)\right)^2 + 4\lambda_1\left(-\lambda_1 \frac{N_1 + N_2}{h_{31}^2}\frac{N_1}{h_{21}^2} + \lambda_3 \frac{N_1 + N_2}{h_{31}^2} + (1-\lambda_3)\frac{N_1}{h_{21}^2}\right).$$

**Corollary:** For the parallel relay channel, the optimization problem is written as [19]

$$R_{\text{full-duplex,DF}} = \max_{P_r, P_s, P_2} \min\{\mathrm{E}\{R_1\}, \mathrm{E}\{R_2\}\},$$

where $R_1(.)$ and $R_2(.)$ are given by

$$R_1 = \sum_{k=1}^{K} C\left(\frac{(P_{rk} + P_{sk})h_{31k}^2 + P_{2k}h_{32k}^2 + 2\sqrt{P_{sk}P_{2k}}h_{31k}h_{32k}}{N_{1k} + N_{2k}}\right)$$

$$R_2 = \sum_{k=1}^{K} C\left(\frac{P_{rk}h_{21k}^2}{N_{1k}}\right).$$

Using the same approach of Theorem 1, the optimum power allocations $P_{rk}$, $P_{sk}$ and $P_{2k}$ for $k = 1, \ldots, K$ are derived for each case that are equal to the optimum power allocations $P_r$, $P_s$ and $P_2$ in the same case of Theorem 1, where $h_{21}$, $h_{31}$, $h_{32}$, $N_1$ and $N_2$ are replaced by $h_{21k}$, $h_{31k}$, $h_{32k}$, $N_{1k}$ and $N_{2k}$, respectively.

## IV. NUMERICAL RESULTS

In this section, some implications of theorems 1 and 2 are discussed for a fading channel. It is assumed that the channel gains, $h_{21}(n)$, $h_{31}(n)$ and $h_{32}(n)$, are independent and identically distributed random variables with Rayleigh probability distribution given by

$$f_h(h) = \frac{h}{0.25}\exp\left(\frac{-h^2}{2 \times 0.25}\right), \forall h \geq 0. \quad (30)$$

In Fig. 2, results of Theorem 2 are demonstrated. Assuming a fixed triple ($\rho_x^2 = 0.2, N_1 = 1, N = 9$)

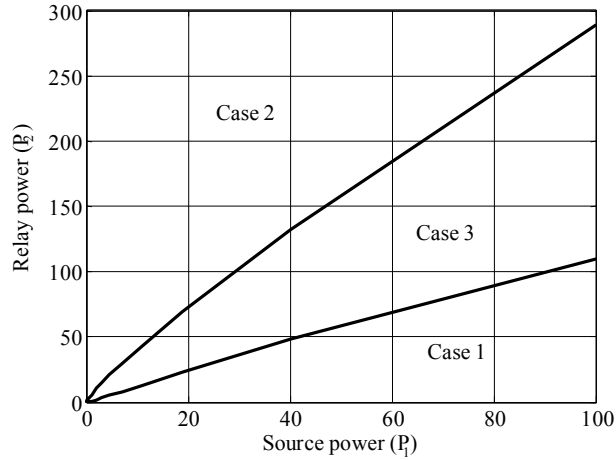

Fig. 2. Three cases of Theorem 2, for fixed triple ($\rho_x^2 = 0.2, N_1 = 1, N = 9$) and various individual power constraints, $\overline{P_1}$ and $\overline{P_2}$.

and for various individual power constraints, $\overline{P_1}$ and $\overline{P_2}$; first, we computed the optimum power allocations, then, three cases are depicted. To evaluate the boundaries between different cases, for each value of $\overline{P_1}$, the following procedure was used.

As $N \gg N_1$, for small $\overline{P_2}$, it is obvious that Case 1 occurs. Therefore, $\overline{P_2}$ is increased and the optimum power allocations are computed for each $\overline{P_2}$. Then, the constraint of Case 1, i.e., $E\{R_2(.)\}_{P_1=P_1^{c_1}} \geq E\{R_1(.)\}_{P_1=P_1^{c_1}, P_2=P_2^{c_1}}$, is checked. Incrementing $\overline{P_2}$ is continued until the constraint of Case 1 fails. This way, the boundary point between cases 1 and 3 is found. Similarly, it is clear that for an extremely large value of $\overline{P_2}$, Case 2 occurs. Thus, $\overline{P_2}$ is decreased and the optimum power allocations are computed and the constraint of Case 2, i.e., $E\{R_2(.)\}_{P_1=P_1^{c_2}} \leq \max_{P_2} E\{R_1(.)\}_{P_1=P_1^{c_2}}$ is checked. Decrementing $\overline{P_2}$ is continued until the constraint of Case 2 fails. This way, the boundary point between cases 2 and 3 is found.

Similarly, in Fig. 3 for fixed pair ($N_1 = 1, N = 2$) and various individual power constraints, $\overline{P_1}$ and $\overline{P_2}$, the optimum $\rho_x$ and the optimum power allocations are evaluated and it is shown that which case occurs. Following the same procedure for plotting Fig. 2, the graphs are plotted. As mentioned in Theorem 1, when the optimization is taken over power allocations and $\rho_x$, Case 1 never happens.

In Fig. 4, a fixed triple ($\overline{P_2} = 1, N_1 = 1, N = 1.6$) was assumed, and for different values of $\overline{P_1}$, the optimum power allocations and the optimum $\rho_x$ are evaluated for each fading state based on the results of Theorem 1, and the optimum rate is plotted. As it is expected, a larger $\overline{P_1}$ (source power), results in a larger rate. It can be seen that for a given data and $\overline{P_1} \leq 1$, Case 2 happens and afterwards Case 3 occurs.

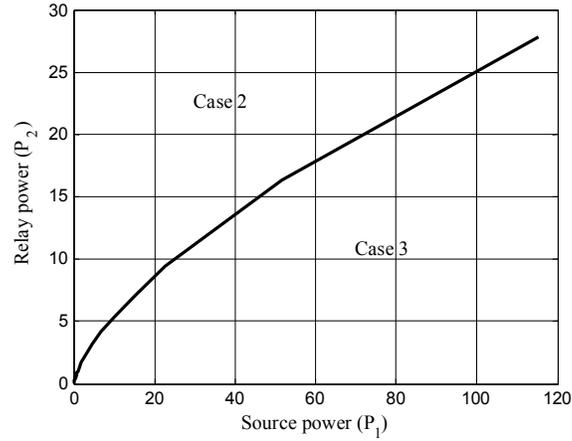

Fig. 3. Different cases of Theorem 1, for fixed triple ($N_1 = 1, N = 2$) and various individual power constraints, $\overline{P_1}$ and $\overline{P_2}$.

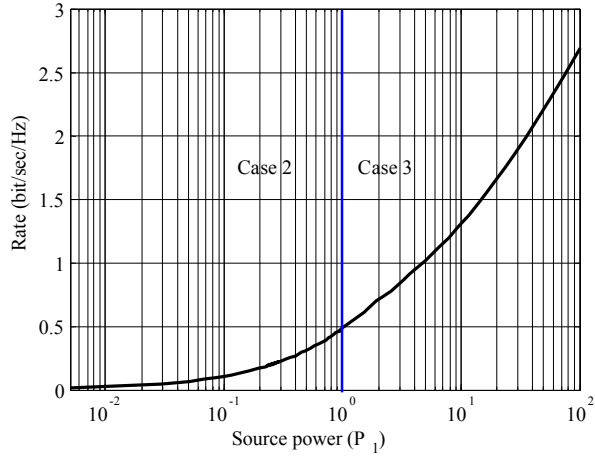

Fig. 4. The optimum rate in the Rayleigh fading environment for parameters ($\overline{P_2} = 1, N_1 = 1, N = 1.6$).

Graph of optimum rate versus different values of relay power constraint, $\overline{P_2}$, for a fixed triple ($\overline{P_1} = 1, N_1 = 1, N = 1.6$) is plotted in Fig. 5. As it is expected, by increasing the relay power constrains, the rate is increased until it reaches a saturated value which shows that the relay has reached the best performance and cannot do better. In other words, the saturated region means that Case 2 happens. In contrast to Fig. 4, for $\overline{P_2} \leq 1$, Case 3 happens and afterwards Case 2 occurs.

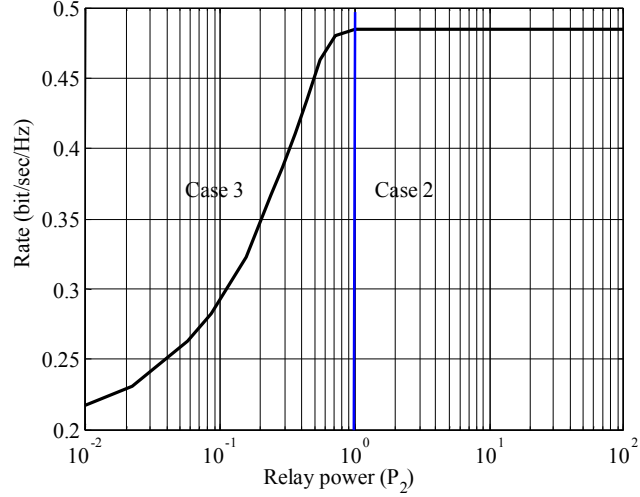

Fig. 5. The optimum achievable rate versus relay power constraint for parameters ($\overline{P_1} = 1, N_1 = 1, N = 1.6$) in the Rayleigh fading environment.

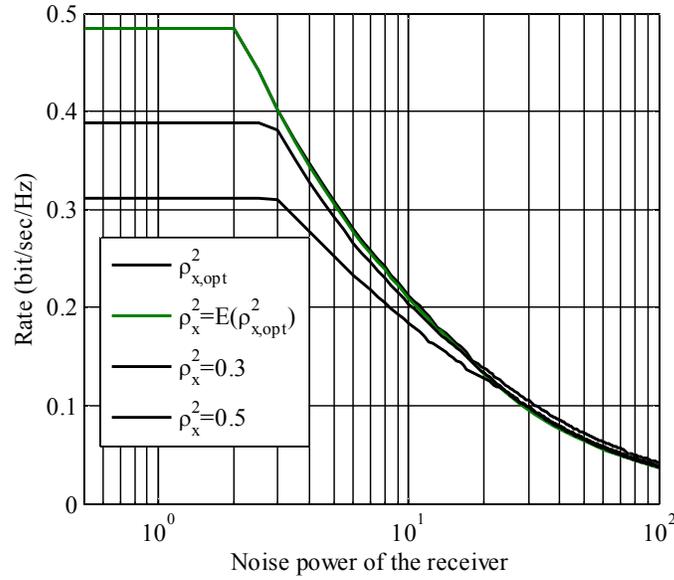

Fig. 6. The optimum achievable rate using Theorems 1 and 2 versus the noise power at the receiver for fixed parameters($\overline{P_1} = 1, \overline{P_1} = 1, N_1 = 1$).

Finally, the results of theorems 1 and 2 are compared for a fixed triple ($\overline{P_1} = 1, \overline{P_1} = 1, N_1 = 1$), in Fig. 6. First, utilizing Theorem 1, optimum power allocations and the optimum $\rho_x$ are derived for each fading state and value of the receiver noise power. Then the optimum rate is plotted versus $N$. Then, utilizing Theorem 2, for two fixed values of $\rho_x = 0.3$ and 0.5, only the optimum power allocations are derived and the rate is plotted versus $N$. It is demonstrated that the optimum rate derived based on Theorem 1 is strictly greater that the rate derived based on the results of Theorem 2. Moreover, for each $N$, we have computed the expected value of optimum $\rho_x$ over different fading processes and used this value in Theorem 2. It is shown that the performance of this scenario is close to the optimum case (results of Theorem 1) and the gap is negligible.

## V. CONCLUSIONS

In this paper, the Gaussian relay channel was considered in which the achievable rate was established for a relay operating in the full-duplex mode, using the decode-and-forward strategy. It was proven that the optimization problem is convex, and the problem can be analyzed analytically. Three distinct cases were considered and for each, optimum power allocations were derived in closed-form. Two scenarios were considered: First, the optimum functions of the correlation coefficient $\rho_x$, between the source and relay powers were derived, and then, to decrease the complexity of the source, assuming a fixed $\rho_x$, the optimum functions of the source and relay power were obtained. Ultimately, implications of our theorems, especially the transitions between the cases and the optimality of the results of Theorem 1 were discussed, in the context of a hypothetical fading channel.

## APPENDIX

***Proof of Lemma 3:*** To maximize the function $\min(\mathrm{E}\{R_1(P_r, P_s, P_2, h_{31}, h_{32})\}, \mathrm{E}\{R_2(P_r, P_s, h_{21})\})$ for all possible positive power allocations $P_1(.)$ for $k = 1,2$, which satisfy two power constraints, three cases should be considered

**Case 1.** If

$$\mathrm{E}\{R_2(.)\}_{P_r=P_r^{c_1}, P_s=P_s^{c_1}} \geq \mathrm{E}\{R_1(.)\}_{P_r=P_r^{c_1}, P_s=P_s^{c_1}, P_2=P_2^{c_1}},$$

then

418

$$\max_{P_r,P_s,P_2} \min(\mathrm{E}\{R_1(.)\}, \mathrm{E}\{R_2(.)\}) \geq \min\left(\mathrm{E}\{R_2(.)\}_{P_r=P_r^{c_1},P_s=P_s^{c_1}}, \mathrm{E}\{R_1(.)\}_{P_r=P_r^{c_1},P_s=P_s^{c_1},P_2=P_2^{c_1}}\right)$$

$$= \mathrm{E}\{R_1(.)\}_{P_r=P_r^{c_1},P_s=P_s^{c_1},P_2=P_2^{c_1}}.$$

Therefore,

$$\max_{P_r,P_s,P_2} \mathrm{E}\{R_1(.)\} = \mathrm{E}\{R_1(.)\}_{P_r=P_r^{c_1},P_s=P_s^{c_1},P_2=P_2^{c_1}}.$$

Thus,

$$\max_{P_r,P_s,P_2} \min(\mathrm{E}\{R_1(.)\}, \mathrm{E}\{R_2(.)\}) = \mathrm{E}\{R_1(.)\}_{P_1=P_1^1,P_2=P_2^1},$$

and $P_r^{c_1}, P_s^{c_1}$ and $P_2^{c_1}$ are the optimum power allocations. The Lagrangian function for deriving the optimum power allocation in this case is

$$\mathcal{L} = \mathrm{E}\{R_1(.)\} + \lambda_1(\overline{P_1} - \mathrm{E}\{P_r + P_s\}) + \lambda_2(\overline{P_2} - \mathrm{E}\{P_2\}). \tag{31}$$

As $\mathrm{E}\{R_1(.)\}$ is convex and the constraints are convex in the space of all power allocations, this optimization problem is convex and has a unique global maximum which is obtained from the Kuhn-Tucker conditions. These conditions are sufficient to derive the maximum and $\lambda_k, k = 1,2$ are uniquely derived from the two power constraints.

**Case 2.** If

$$\mathrm{E}\{R_2(.)\}_{P_r=P_r^{c_2},P_s=P_s^{c_2}} \leq \max_{P_2} \mathrm{E}\{R_1(.)\}_{P_r=P_r^{c_2},P_s=P_s^{c_2}},$$

then

$$\min\left(\mathrm{E}\{R_2(.)\}_{P_r=P_r^{c_2},P_s=P_s^{c_2}}, \max_{P_2} \mathrm{E}\{R_1(.)\}_{P_r=P_r^{c_2},P_s=P_s^{c_2}}\right) = \mathrm{E}\{R_2(.)\}_{P_r=P_r^{c_2},P_s=P_s^{c_2}},$$

where $P_2 = P_2^{c_2}$ is the power allocation which maximizes $\mathrm{E}\{R_1(.)\}_{P_r=P_r^{c_2},P_s=P_s^{c_2}}$. Thus,

$$\max_{P_r,P_s,P_2} \min(\mathrm{E}\{R_1(.)\}, \mathrm{E}\{R_2(.)\}) \geq \min\left(\mathrm{E}\{R_2(.)\}_{P_r=P_r^{c_2},P_s=P_s^{c_2}}, \mathrm{E}\{R_1(.)\}_{P_r=P_r^{c_2},P_s=P_s^{c_2},P_2=P_2^{c_2}}\right)$$

$$= \mathrm{E}\{R_2(.)\}_{P_r=P_r^{c_2},P_s=P_s^{c_2}},$$

and since $\max_{P_r,P_s} \mathrm{E}\{R_2(.)\} = \mathrm{E}\{R_2(.)\}_{P_r=P_r^{c_2},P_s=P_s^{c_2}}$, we have

$$\max_{P_r,P_s,P_2} \min(\mathrm{E}\{R_1(.)\}, \mathrm{E}\{R_2(.)\}) = \mathrm{E}\{R_2(.)\}_{P_r=P_r^{c_2}, P_s=P_s^{c_2}}.$$

Thus, $P_r^{c_2}$, $P_s^{c_2}$ and $P_2^{c_2}$ are optimum power allocations. Accordingly, the Lagrangian function for deriving the optimum power allocation in this case would be as:

$$\mathcal{L} = \mathrm{E}\{R_2(.)\} + \lambda_1(\overline{P_1} - \mathrm{E}\{P_r + P_s\}). \tag{32}$$

As $\mathrm{E}\{R_2(.)\}$ is convex and the constraints are convex in the space of all power allocations, this optimization problem is convex and has a unique global maximum which is obtained from the Kuhn-Tucker conditions. These conditions are sufficient to derive the maximum, and $\lambda_1$ is uniquely derived from the power constraint.

**Case 3.** By contradiction, we assume that the optimum solution: $\max_{P_r,P_s,P_2} \min(\mathrm{E}\{R_1(.)\}, \mathrm{E}\{R_2(.)\})$ does not occur on the surface of $J$, and one of the functions $\mathrm{E}\{R_1(.)\}$ and $\mathrm{E}\{R_2(.)\}$ are larger than the other. Without loss of generality, assume

$$\mathrm{E}\{R_2(.)\}_{P_r=P_r^*, P_s=P_s^*} > \mathrm{E}\{R_1(.)\}_{P_r=P_r^*, P_s=P_s^*, P_1=P_2^*},$$

where $P_r^*$, $P_s^*$ and $P_2^*$ are the optimum solution of $\max_{P_r,P_s,P_2} \min(\mathrm{E}\{R_1(.)\}, \mathrm{E}\{R_2(.)\})$. Thus, as the function $\mathrm{E}\{R_2(.)\}$ is a continuous function, and we have

$$\mathrm{E}\{R_2(.)\}_{P_r=P_r^*, P_s=P_s^*} > \mathrm{E}\{R_1(.)\}_{P_r=P_r^*, P_s=P_s^*, P_1=P_2^*}, \text{ and } \mathrm{E}\{R_2(.)\}_{P_r=P_r^{c_2}, P_s=P_s^{c_2}}$$
$$\leq \mathrm{E}\{R_1(.)\}_{P_r=P_r^{c_2}, P_s=P_s^{c_2}, P_2=P_2^{c_2}},$$

there exists a real number $0 \leq \alpha < 1$ such that

$$\mathrm{E}\{R_2\}_{P_r=\alpha P_r^*+(1-\alpha)P_r^{c_2}, P_s=\alpha P_s^*+(1-\alpha)P_s^{c_2}}$$
$$= \mathrm{E}\{R_1\}_{P_r=\alpha P_r^*+(1-\alpha)P_r^{c_2}, P_s=\alpha P_s^*+(1-\alpha)P_s^{c_2}, P_2=\alpha P_2^*+(1-\alpha)P_2^{c_2}}. \tag{33}$$

On the other hand, as the functions $\mathrm{E}\{R_1(.)\}$ and $\mathrm{E}\{R_2(.)\}$ are convex functions, we have

$$\mathrm{E}\{R_2\}_{P_r=\alpha P_r^*+(1-\alpha)P_r^{c_2}, P_s=\alpha P_s^*+(1-\alpha)P_s^{c_2}} \geq \alpha \mathrm{E}\{R_2\}_{P_r=P_r^*, P_s=P_s^*} + (1-\alpha)\mathrm{E}\{R_2\}_{P_r=P_r^{c_2}, P_s=P_s^{c_2}}$$
$$\geq \mathrm{E}\{R_2\}_{P_r=P_r^*, P_s=P_s^*}. \tag{34}$$

Therefore, from equations (33) and (34), we obtain

$$\min \left( E\{R_2\}_{P_r=\alpha P_r^*+(1-\alpha)P_r^{c_2}, P_s=\alpha P_s^*+(1-\alpha)P_s^{c_2}}, E\{R_1\}_{P_r=\alpha P_r^*+(1-\alpha)P_r^{c_2}, P_s=\alpha P_s^*+(1-\alpha)P_s^{c_2}, P_2=\alpha P_2^*+(1-\alpha)P_2^{c_2}} \right) \quad (35)$$

$$\geq E\{R_2\}_{P_r=P_r^*, P_s=P_s^*} = \min\left( E\{R_2\}_{P_r=P_r^*, P_s=P_s^*}, E\{R_1\}_{P_r=P_r^*, P_s=P_s^*, P_2=P_2^*} \right)$$

Hence, $P_r = \alpha P_r^* + (1-\alpha)P_r^{c_2}$, $P_s = \alpha P_s^* + (1-\alpha)P_s^{c_2}$ and $P_2 = \alpha P_2^* + (1-\alpha)P_2^{c_2}$ lead to the larger values than our presumed optimum solution; $P_r^*, P_s^*, P_2^*$. Similar discussion for the case $E\{R_2\}_{P_r=P_r^*, P_s=P_s^*} < E\{R_1\}_{P_r=P_r^*, P_s=P_s^*, P_2=P_2^*}$ can be given. Therefore, the optimization problem can be rewritten as

$$\max_{P_1, P_2} E\{R_1\} \quad (36)$$
$$\text{s.t. } E\{R_1\} = E\{R_2\}, E\{P_r + P_s\} \leq \overline{P_1}, E\{P_2\} \leq \overline{P_2}.$$

Thus, the Lagrangian function for deriving this optimum power allocation is given by

$$\mathcal{L} = E\{R_1(.)\} + \lambda_3(E\{R_2(.)\} - E\{R_1(.)\}) + \lambda_1(\overline{P_1} - E\{P_r + P_s\}) + \lambda_2(\overline{P_2} - E\{P_2\}). \quad (37)$$

As this case is a special case of the convex optimization problem (5), it is clear that the optimization problem (37) and the objective function $\lambda_3 E\{R_1(P_r, P_s, P_2, h_{31}, h_{32})\} + (1-\lambda_3)E\{R_2(P_r, P_s, h_{21})\}$ are convex and $0 \leq \lambda_3 \leq 1$.

This optimization problem has a unique global maximum which is obtained from the Kuhn-Tucker conditions, and these conditions are sufficient to derive the maximum. $\lambda_j, j = 1,2,3$ are unique and derived from the two power constraints and the equality $E\{R_1\} = E\{R_2\}$. ∎

It is worth noting that, the optimization problem $\max_{P_1 P_2} \min(E\{R_1\}, E\{R_2\})$ can be solved from the optimization problem $\min_{0\leq\lambda_i\leq 1, \lambda_1+\lambda_2=1} \max_{P_1,P_2} \sum_{i=1}^{2} \lambda_i E\{R_i\}$ similar to [21], however utilizing Lemma 3, the optimization problem becomes much simpler and there is no need to minimize the problem over $\lambda_i$. Since in Lemma 2, the optimization problem is convex and has only one maximum, the $\lambda_i$ are unique and the Kuhn-Tucker conditions are sufficient for evaluating $\lambda_i$.

***Proof of Theorem 1:*** In the following, the optimum power allocations, $P_s(.)$, $P_r(.)$, and $P_2(.)$ are derived for the three cases discussed in Lemma 3.

**Case 1:** If

$$E\{R_2(P_r, P_s, h_{21})\}_{P_r=P_r^{c_1}, P_s=P_s^{c_1}} \geq E\{R_1(P_r, P_s, P_2, h_{31}, h_{32})\}_{P_r=P_r^1, P_s=P_s^1, P_2=P_2^1},$$

then

$$\max_{P_r, P_{s_1}, P_2} \min(E\{R_1(.)\}, E\{R_2(.)\}) = E\{R_1(.)\}_{P_r=P_r^{c_1}, P_s=P_s^{c_1}, P_2^{c_1}=P_2^{c_1}},$$

and the optimum power allocations are obtained from the following optimization problem

$$\max_{P_r, P_s, P_2} E\{R_1\} \tag{38}$$

$$\text{s.t.} \quad E\{P_r + P_s\} \leq \bar{P}_1, E\{P_2\} \leq \bar{P}_2. \tag{39}$$

Therefore, the Lagrangian function is

$$\mathcal{L} = E\{R_1\} + \lambda_1(\bar{P}_1 - E\{P_r + P_s\}) + \lambda_2(\bar{P}_2 - E\{P_2\}). \tag{40}$$

It is obvious that If the total power of user 1 is lonely to be allocated to $P_s$, i.e., $P_r = 0, P_s = P_1$, higher rate is achieved. Therefore, derivative of $\mathcal{L}$ regardless of the expectation, with respect to $P_s$ and $P_2$, we obtain

$$\lambda_k = \frac{h_{3k}^2 + \sqrt{\frac{P_{3-k}}{P_k}} h_{31} h_{32}}{N_1 + N_2 + P_1 h_{31}^2 + P_2 h_{32}^2 + 2\sqrt{P_1 P_2} h_{31} h_{32}}. \tag{41}$$

Dividing the equations (41) for $k = 1$ by (41) for $k = 2$, and using $\Gamma = \sqrt{P_2/P_1}$, $\Gamma$ is derived.

$$\Gamma = \frac{\lambda_1 h_{32}}{\lambda_2 h_{31}}. \tag{42}$$

Substituting $\sqrt{P_2} = \Gamma \sqrt{P_1}$ in (41) for $k = 1$, $\lambda_1$ is obtained.

$$\lambda_1 = \frac{h_{31}^2 + \Gamma h_{31} h_{32}}{N_1 + N_2 + P_1(h_{31}^2 + \Gamma^2 h_{32}^2 + 2\Gamma h_{31} h_{32})}. \tag{43}$$

Thus, the power allocations $P_1(.)$ and $P_2(.)$ are derived from (43) as

$$P_1^{c_1}(h_{31}, h_{32}, \Gamma) = \left(\frac{\frac{h_{31}^2 + \Gamma h_{31} h_{32}}{\lambda_1} - N_1 - N_2}{h_{31}^2 + \Gamma^2 h_{32}^2 + 2\Gamma h_{31} h_{32}}\right)^+ \quad (44)$$

$$P_2^{c_1}(h_{31}, h_{32}, \Gamma) = \Gamma^2 P_s^1(h_{31}, h_{32}, \lambda, \Gamma),$$

where power levels $\lambda_k, k = 1,2$ are unique and obtained from the two power constraints (39).

**Case 2.** If

$$E\{R_2(.)\}_{P_r = P_r^{c_2}, P_s = P_s^{c_2}} \leq \max_{P_2} E\{R_1(.)\}_{P_r = P_r^{c_2}, P_s = P_s^{c_2}},$$

then

$$\max_{P_r, P_{s_1}, P_2} \min(E\{R_1(.)\}, E\{R_2(.)\}) = E\{R_2(.)\}_{P_r = P_r^{c_2}, P_s = P_s^{c_2}, P_2 = P_2^{c_2}},$$

and the optimum power allocations are obtained from the following optimization problem

$$\max_{P_r, P_s, P_2} E\{R_2\} \quad (45)$$

$$\text{s.t.} \quad E\{P_r + P_s\} \leq \overline{P_1}, E\{P_2\} \leq \overline{P}_2. \quad (46)$$

From (7), it is clear that $E\{R_2\}$ is independent of $P_2$. As a result, the constraint with respect to $P_2$ can be omitted. It is also obvious that setting $P_s = 0$, rate is increased. Therefore, the Lagrangian function is

$$\mathcal{L} = E\{R_2\} + \lambda_1(\overline{P_1} - E\{P_r\}). \quad (47)$$

Derivative of $\mathcal{L}$ regardless of the expectation with respect to $P_r$, we have

$$P_r^{c_2}(h_{21}) = \left(\frac{1}{2\lambda_1} - \frac{N_1}{h_{21}^2}\right)^+, \quad (48)$$

where $\lambda_1$ is obtained from the power constraint (46).

From Lemma 3-Case 2, comparing $E\{R_2(.)\}_{P_r = P_r^{c_2}, P_s = 0}$ and $\max_{P_2} E\{R_1(.)\}_{P_r = P_r^{c_2}, P_s = 0}$, is needed. Thus, following optimization problem must be solved

$$\max_{P_2} E\{R_1\}_{P_r = P_r^{c_2}, P_s = 0} \quad : \quad \text{s.t.} \quad E\{P_2\} \leq \overline{P}_2, \quad (49)$$

where $P_r^{c_2}$ is evaluated in (48). Then, the Lagrangian function is written as

$$\mathcal{L} = \mathrm{E}\{R_1\}_{P_r = P_r^2, P_s = P_s^2} + \lambda_1(\bar{P_1} - \mathrm{E}\{P_1\}). \tag{50}$$

Regardless of expectation and derivative of $\mathcal{L}$ with respect to $P_2$, $\lambda_1$ is

$$\lambda_1 = \frac{h_{32}^2}{N_1 + N_2 + P_r^2 h_{31}^2 + P_2 h_{32}^2}, \tag{51}$$

and $P_2^{c_2}(.)$ is obtained as

$$P_2^{c_2}(h_{21}, h_{31}, h_{32}) = \left(\frac{1}{\lambda_1} - \frac{N_1 + N_2}{h_{32}^2} - \frac{P_r^2 h_{31}^2}{h_{32}^2}\right)^+, \tag{52}$$

where $\lambda_2$ is obtained from the power constraint (46).

**Case 3**. If none of the Cases 1 and 2 occurs, then $\max_{P_r, P_s, P_2} \min(\mathrm{E}\{R_1(.)\}, \mathrm{E}\{R_2(.)\})$ occurs on the surface $J$ and the optimum power allocations are obtained from the following optimization problem

$$\begin{aligned}&\max_{P_r, P_s, P_2} \mathrm{E}\{R_1\} \\ &\text{s.t.} \quad \mathrm{E}\{P_r + P_s\} \leq \bar{P_1}, \mathrm{E}\{P_2\} \leq \bar{P_2}, \mathrm{E}\{R_1\} = \mathrm{E}\{R_2\}.\end{aligned} \tag{53}$$

Therefore, the Lagrangian function is formulated as

$$\mathcal{L} = \mathrm{E}\{R_1\} + \lambda_3(\mathrm{E}\{R_2\} - \mathrm{E}\{R_1\}) + \lambda_1(\bar{P_1} - \mathrm{E}\{P_r + P_s\}) + \lambda_2(\bar{P_2} - \mathrm{E}\{P_2\}). \tag{54}$$

Derivative of $\mathcal{L}$ regardless of expectation, with respect to $P_r$, $P_s$ and $P_2$, we have

$$\lambda_1 = \frac{\lambda h_{21}^2}{N_1 + P_r h_{21}^2} + \frac{(1-\lambda) h_{31}^2}{N_1 + N_2 + (P_r + P_s) h_{31}^2 + P_2 h_{32}^2 + 2\sqrt{P_s P_2} h_{31} h_{32}} \tag{55}$$

$$\lambda_2 = \frac{(1-\lambda)\left(h_{32}^2 + \sqrt{\frac{P_s}{P_2}} h_{31} h_{32}\right)}{N_1 + N_2 + (P_r + P_s) h_{31}^2 + P_2 h_{32}^2 + 2\sqrt{P_s P_2} h_{31} h_{32}} \tag{56}$$

$$\lambda_1 = \frac{(1-\lambda)\left(h_{31}^2 + \sqrt{\frac{P_2}{P_s}} h_{31} h_{32}\right)}{N_1 + N_2 + (P_r + P_s) h_{31}^2 + P_2 h_{32}^2 + 2\sqrt{P_s P_2} h_{31} h_{32}}. \tag{57}$$

Dividing (56) by (57) and using $\Gamma = \sqrt{P_2/P_s}$, $\Gamma$ is given by

$$\Gamma = \frac{\lambda_1 h_{32}}{\lambda_2 h_{31}}. \tag{58}$$

Employing $\Gamma$ in equations (55) and (57), we arrive at

$$\lambda_1 = \frac{\lambda h_{21}^2}{N_1 + P_r h_{21}^2} + \frac{(1-\lambda)h_{31}^2}{N_1 + N_2 + P_r h_{31}^2 + P_s(h_{31}^2 + \Gamma^2 h_{32}^2 + 2\Gamma h_{31} h_{32})} \tag{59}$$

$$\lambda_1 = \frac{(1-\lambda)(h_{31}^2 + \Gamma h_{31} h_{32})}{N_1 + N_2 + P_r h_{31}^2 + P_s(h_{31}^2 + \Gamma^2 h_{32}^2 + 2\Gamma h_{31} h_{32})}. \tag{60}$$

Substituting $N_1 + N_2 + P_r h_{31}^2 + P_s(h_{31}^2 + \Gamma^2 h_{32}^2 + 2\Gamma h_{31} h_{32})$ from (60) in (59), $P_r(.)$ is derived as

$$P_r^{c_3}(h_{21}, h_{31}, h_{32}) = \left(\frac{\lambda(h_{31}^2 + \Gamma h_{31} h_{32})}{\lambda_1 \Gamma h_{31} h_{32}} - \frac{N_1}{h_{21}^2}\right)^+. \tag{61}$$

Substituting $P_r^{c_3}(.)$ in (60), $P_s(.)$ is computed as

$$P_s^{c_3}(h_{21}, h_{31}, h_{32}) = \frac{\left(\frac{(1-\lambda)(h_{31}^2 + \Gamma h_{31} h_{32})}{\lambda_1} - N_1 - N_2 - P_r h_{31}^2\right)^+}{h_{31}^2 + \Gamma^2 h_{32}^2 + 2\Gamma h_{31} h_{32}} \tag{62}$$

$$P_2^{c_3}(h_{21}, h_{31}, h_{32}) = \Gamma^2 P_s^{c_3}(h_{21}, h_{31}, h_{32}).$$

The power levels $\lambda_j, j = 1,2,3$ are derived from the two power constraints, and the equality $E\{R_1\} = E\{R_2\}$. ∎

***Proof of Theorem 2:*** In the following, the optimum power allocations, $P_k(.), k = 1,2$ are derived for the three cases discussed in Lemma 3, similar to Theorem 1.

**Case 1:** If

$$E\{R_2(P_1, \rho_x, h_{21})\}_{P_1 = P_1^{c_1}} E\{R_1(P_1, P_2, \rho_x, h_{31}, h_{32})\}_{P_1 = P_1^{c_1}, P_2 = P_2^{c_1}},$$

then

$$\max_{P_1, P_2} \min(E\{R_1(.)\}, E\{R_2(.)\}) = E\{R_1(.)\}_{P_1 = P_1^{c_1}, P_2 = P_2^{c_1}}$$

and the optimum power allocations are obtained from the following optimization problem:

$$\max_{P_1, P_2} E\{R_1\} \tag{63}$$

$$\text{s.t.} \quad \text{E}\{P_k\} \leq \overline{P_k}, \quad \forall k = 1,2. \tag{64}$$

Therefore, the Lagrangian function is formulated as

$$\mathcal{L} = \text{E}\{R_1\} + \lambda_1(\overline{P_1} - \text{E}\{P_1\}) + \lambda_2(\overline{P_2} - \text{E}\{P_2\}). \tag{65}$$

Utilizing Lemma 1 and derivative of $\mathcal{L}$ with respect to $P_k$ for $k = 1,2$ regardless of expectation, we arrive at

$$\lambda_k = \frac{h_{3k}^2 + \rho_x\sqrt{\frac{P_{3-k}}{P_k}}h_{31}h_{32}}{N_1 + N_2 + P_1 h_{31}^2 + P_2 h_{32}^2 + 2\rho_x\sqrt{P_1 P_2}h_{31}h_{32}}. \tag{66}$$

Dividing (66) for $k = 1$ by (66) for $k = 2$, we obtain

$$\frac{\lambda_1}{\lambda_2} = \frac{h_{31}^2 + \rho_x\sqrt{\frac{P_2}{P_1}}h_{31}h_{32}}{h_{32}^2 + \rho_x\sqrt{\frac{P_1}{P_2}}h_{31}h_{32}}. \tag{67}$$

Defining $\Gamma \coloneqq \sqrt{P_2/P_1}$ and substituting in (67), $\Gamma$ is derived from a quadratic equation. So, $\Gamma$ is evaluated as

$$\Gamma = \frac{-\lambda_2 h_{31}^2 + \lambda_1 h_{32}^2 + \sqrt{(\lambda_2 h_{31}^2 - \lambda_1 h_{32}^2)^2 + 4\lambda_1\lambda_2(\rho_x h_{31}h_{32})^2}}{2\lambda_2 \rho_x h_{31}h_{32}}. \tag{68}$$

Substituting $\sqrt{P_2} = \Gamma\sqrt{P_1}$ in (66) for $k = 1$, $\lambda_1$ is derived.

$$\lambda_1 = \frac{h_{31}^2 + \rho_x \Gamma h_{31}h_{32}}{N_1 + N_2 + P_1(h_{31}^2 + \Gamma^2 h_{32}^2 + 2\rho_x \Gamma h_{31}h_{32})}. \tag{69}$$

Thus, the power allocations $P_k(.)$ for $k = 1,2$ are derived from (69) as

$$P_1^{c_1}(h_{31}, h_{32}, \Gamma) = \left(\frac{\frac{h_{31}^2 + \rho_x \Gamma h_{31}h_{32}}{\lambda_1} - N_1 - N_2}{h_{31}^2 + \Gamma^2 h_{32}^2 + 2\rho_x \Gamma h_{31}h_{32}}\right)^+ \tag{70}$$

$$P_2^{c_1}(h_{31}, h_{32}, \Gamma) = \Gamma^2 P_1^{c_1}(h_{31}, h_{32}, \lambda, \Gamma).$$

The power levels $\lambda_k, k = 1,2$ are unique and obtained from the two power constraints (64).

**Case 2:** If

$$E\{R_2(.)\}_{P_1=P_1^{c_2}} \leq \max_{P_2} E\{R_1(.)\}_{P_1=P_1^{c_2}},$$

then

$$\max_{P_1,P_2} \min(E\{R_1(.)\}, E\{R_2(.)\}) = E\{R_2(.)\}_{P_1=P_1^{c_2}}$$

and the optimum power allocations are obtained from the following optimization problem;

$$\max_{P_1,P_2} E\{R_2\} \qquad (71)$$

$$\text{s.t.} \quad E\{P_k\} \leq \overline{P_k}, \quad \forall k = 1,2. \qquad (72)$$

From (10), it is clear that $E\{R_2\}$ is independent of $P_2$. As a result, the constraint with respect to $P_2$ can be omitted and the Lagrangian function is

$$\mathcal{L} = E\{R_2\} + \lambda(\overline{P_1} - E\{P_1\}). \qquad (73)$$

Derivative of $\mathcal{L}$ with respect to $P_1$ regardless of expectation, we arrive at

$$P_1^{c_2}(h_{21}) = \left(\frac{1}{2\lambda_1} - \frac{N_1}{(1-\rho_x^2)h_{21}^2}\right)^+. \qquad (74)$$

From Lemma 3-Case 2, comparing $E\{R_2(P_1, \rho_x, h_{21})\}_{P_1=P_1^{c_2}}$ and $\max_{P_2} E\{R_1(P_1, P_2, \rho_x, h_{31}, h_{32})\}_{P_1=P_1^{c_2}}$, is needed. Thus, following optimization problem is solved

$$\max_{P_2} E\{R_1\}_{P_1=P_1^{c_2}} : \text{s.t.} \quad E\{P_2\} \leq \overline{P_2} \qquad (75)$$

where $P_1^{c_2}$ is evaluated in (74). Then, the Lagrangian function is written as

$$\mathcal{L} = E\{R_1\}_{P_1=P_1^2} + \lambda_2(\overline{P_2} - E\{P_2\}). \qquad (76)$$

Derivative of $\mathcal{L}$ with respect to $P_2$ regardless of expectation, (77) is derived.

$$\lambda_2 = \frac{h_{32}^2 + \rho_x \sqrt{\frac{P_1^2}{P_2}} h_{31}h_{32}}{N_1 + N_2 + P_1^2 h_{31}^2 + P_2 h_{32}^2 + 2\rho_x \sqrt{P_1^2 P_2} h_{31}h_{32}}. \qquad (77)$$

Collecting terms of $\left(\sqrt{P_2}\right)^m, m = 1,2,3$ in (77), a cubic equation is derived based on the terms $\sqrt{P_2}$. Solving the cubic equation, $P_2(.)$ is derived as

$$P_2^{c_2}(h_{21}, h_{31}, h_{32})$$

$$= \frac{1}{h_{32}^2}\left(\left(\frac{-F}{3} - \frac{1}{3}\sqrt[3]{\frac{1}{2}\left(A + \sqrt{(A^2 - 4(F^2 - 3G)^3)^+}\right)}\right.\right. \tag{78}$$

$$\left.\left.- \frac{1}{3}\sqrt[3]{\frac{1}{2}\left(A - \sqrt{(A^2 - 4(F^2 - 3G)^3)^+}\right)}\right)^+\right)^2$$

where $A$, $F$ and $G$ are defined as

$$A \coloneqq 2F^3 - 9FG - 27\frac{h_{32}^2}{2\lambda_2}\rho_x h_{31}\sqrt{P_1^2}$$

$$F \coloneqq 2\rho_x h_{31}\sqrt{P_1^2} \tag{79}$$

$$G \coloneqq N_1 + N_2 - \frac{h_{32}^2}{2\lambda_2}$$

and $\lambda_2$ is obtained from the power constraint (72).

**Case 3.** If none of the Cases 1 and 2 occurs, then $\max_{P_1,P_2} \min(E\{R_1(.)\}, E\{R_2(.)\})$ occurs on the surface $J$ and the optimum power allocations are obtained from the following optimization problem;

$$\max_{P_1,P_2} E\{R_1\} \tag{80}$$

$$\text{s.t.} \quad E\{P_k\} \leq \overline{P_k}, \forall k = 1,2, E\{R_1\} = E\{R_2\}. \tag{81}$$

Therefore, the Lagrangian function is formulated as

$$\mathcal{L} = E\{R_1\} + \lambda_3(E\{R_2\} - E\{R_1\}) + \lambda_1(\overline{P_1} - E\{P_1\}) + \lambda_2(\overline{P_2} - E\{P_2\}). \tag{82}$$

Derivative of $\mathcal{L}$ with respect to $P_k, k = 1,2$, regardless of expectation, we get

$$\lambda_1 = \frac{\lambda_3(1 - \rho_x^2)h_{21}^2}{N_1 + (1 - \rho_x^2)P_1 h_{21}^2} + \frac{(1 - \lambda_3)\left(h_{31}^2 + \rho_x\sqrt{\frac{P_2}{P_1}}h_{31}h_{32}\right)}{N_1 + N_2 + P_1 h_{31}^2 + P_2 h_{32}^2 + 2\rho_x\sqrt{P_1 P_2}h_{31}h_{32}} \tag{83}$$

$$\lambda_2 = \frac{(1-\lambda_3)\left(h_{32}^2 + \rho_x\sqrt{\frac{P_1}{P_2}}h_{31}h_{32}\right)}{N_1 + N_2 + P_1 h_{31}^2 + P_2 h_{32}^2 + 2\rho_x\sqrt{P_1 P_2}h_{31}h_{32}}. \tag{84}$$

Using $\Gamma = \sqrt{P_2/P_1}$ and substituting $N_1 + N_2 + P_1 h_{31}^2 + P_2 h_{32}^2 + 2\rho_x\sqrt{P_1 P_2}h_{31}h_{32}$ from (84) in (83), we obtain

$$\frac{\lambda_3(1-\rho_x^2)h_{21}^2}{N_1 + (1-\rho_x^2)P_1 h_{21}^2} = \lambda_1 - \frac{\lambda_2(h_{31}^2 + \rho_x\Gamma h_{31}h_{32})}{h_{32}^2 + \frac{\rho_x}{\Gamma}h_{31}h_{32}}. \tag{85}$$

Thus, solving (85), $P_k(.), k = 1,2$ are derived as

$$P_1^{c_3}(\lambda, \lambda_1, \lambda_2, h_{21}, h_{31}, h_{32}, \Gamma) = \left(\frac{\lambda}{\lambda_1 - \frac{\lambda_2(h_{31}^2 + \rho_x\Gamma h_{31}h_{32})}{h_{32}^2 + \frac{\rho_x}{\Gamma}h_{31}h_{32}}} - \frac{N_1}{(1-\rho_x^2)h_{21}^2}\right)^+ \tag{86}$$

$$P_2^{c_3}(\lambda, \lambda_1, \lambda_2, h_{21}, h_{31}, h_{32}, \Gamma) = \Gamma^2 P_1^{c_3}(\lambda_1, \lambda_2, \lambda_3, h_{21}, h_{31}, h_{32}, \Gamma).$$

Substituting $P_k^{c_3}(.), k = 1,2$ in (84), we arrive at

$$\frac{(1-\lambda_3)}{\lambda_2}\left(h_{32}^2 + \frac{\rho_x h_{31}h_{32}}{\Gamma}\right) = N_1 + N_2 + (h_{31}^2 + \Gamma^2 h_{32}^2 + 2\rho_x\Gamma h_{31}h_{32})$$
$$\times \left(\frac{\lambda_3}{\lambda_1 - \frac{\lambda_2(h_{31}^2 + \rho_x\Gamma h_{31}h_{32})}{h_{32}^2 + \frac{\rho_x}{\Gamma}h_{31}h_{32}}} - \frac{N_1}{(1-\rho_x^2)h_{21}^2}\right). \tag{87}$$

It is clear that (87) is a quintic equation with respect to $\Gamma$, and $\Gamma$ could be obtained numerically for each state.

The power levels $\lambda_j, j = 1,2,3$ are derived from the two power constraints and the equality $E\{R_1\} = E\{R_2\}$ in (81). ∎

## REFERENCES

[1] E. C. van der Muelen, *Transmission of information in a T-terminal discrete memoryless channel.* Ph.D. dissertation, Dep. Of Statist., Univ. Calif., Berkeley, 1968.

[2] ----------, "Three-terminal communication channels," *Adv. Appl. Probab.*, vol. 3, pp. 120–154, 1971.

[3] ----------, "A survey of multi-way channels in information theory:1961–1976," *IEEE Trans. Inform. Theory*, vol. 23, no. 2, pp. 1–37, Jan.1977.

[4] T. M. Cover and A. El Gamal, "Capacity theorems for the relay channel," *IEEE Trans. Inform. Theory*, vol. 25, no.5, pp. 572–584, Sept. 1979.

[5] M. R. Aref, *Information Flow in Relay Networks*. PhD dissertation, Stanford Univ., Stanford, CA, Oct., 1980.

[6] S. Boyd and L. Vandenberghe, *Convex Optimization*. Cambridge University Press, 2004.

[7] W. Yu, W. Rhee, S. Boyd, and J. M. Cioffi, "Iterative water-filling for Gaussian vector multiple-access channels," *IEEE Trans. Inform. Theory*, vol. 50, no. 1, pp. 145–152, Jan. 2004.

[8] D.P. Bertsekas, A. Nedic, and A.E. Ozdaglar, *Convex Analysis and Optimization*. Athena Scientific. Cambridge, Massachusetts, 2003.

[9] A. ParandehGheibi, A. Eryilmaz, A. Ozdaglar, and M. Medard, "Rate and power allocation in fading multiple access channels," *6th Int. Symp.on Modeling and Optimization in Mobile, Ad Hoc, and Wireless Networks and Workshops (WiOPT 2008),* Berlin, pp. 282-287, Apr., 2008.

[10] S. Vishwanath, S.A. Jafar, and A. Goldsmith, "Optimum power and rate allocation strategies for multiple access fading channels," in *Proc. Vehicular Technology Conf., (VTC 2001)*, pp.2888-2892, vol. 4, May 2001.

[11] I. E. Telatar, "Capacity of multi-antenna Gaussian channels," *European Trans. Telecommun.*, vol. 10(6), pp. 585–596, Nov. 1999.

[12] A. Høst-Madsen and J. Zhang, "Capacity bounds and power allocation for wireless relay channels," *IEEE Trans. Inf. Theory*, vol. 51, no. 6, pp. 2020–2040, June. 2005.

[13] Y. Yao, X. Cai, and G. B. Giannakis, "On energy efficiency and optimum resource allocation of relay transmissions in the low-power regime," *IEEE Trans. Wireless Commun.*, vol. 4, no. 6, pp. 2917–2927,Nov. 2005.

[14] I. Maric and R. Yates, "Forwarding strategies for Gaussian parallel-relay networks," in *Proc. IEEE Int. Symp. Information Theory* , Chicago, IL, Jun./Jul. 2004, p. 269.

[15] D. Gunduz and E. Erkip, "Opportunistic cooperation by dynamic resource allocation," *IEEE Trans. Wireless Commun.*, vol. 6, no. 4, pp. 1446–1454, Apr. 2007.

[16] M. EI Soussi, A. Zaidi, J. Louveaux and L. Vandendorpe, "Sum-rate optimized power allocation for the OFDM multiple access relay channel," in *Proc. 3th Int. Symp. Commun., Control and Signal Processing (ISCCSP),* Limassol, Cyprus, pp.1-6, Mar. 2010.

[17] K.T. Phan, Tho Le-Ngoc, S. A. Vorobyov, C. Tellambura, "Power allocation in wireless multi-user relay networks,"*IEEE Trans. Wireless Comm*, vol. 8, no. 5, pp. 2535-2545, May 2009.

[18] M. Pischella and D. Le Ruyet, "Optimal Power Allocation for the Two-Way Relay Channel with Data Rate Fairness," *IEEE, Commun Letters*, vol. 15, no. 9, pp. 959-961, Sep. 2011.

[19] Y. Liang, V. Veeravalli, and H. Vincent Poor. "Resource Allocation for Wireless Fading Relay Channels: Max-Min Solution," *IEEE Trans. Inform. Theory*, vol. 53, no. 10, pp.3432-3453, October2007.

[20] L. Zhang, J. Jiang, A. J. Goldsmith, and S. Cui, "study of Gaussian Relay Channels with correlated noises," *IEEE Trans. Commun. Theory*, vol. 59, no. 3, pp. 863-876, March 2011.

[21] Y. Geng, A. Gohari, C. Nair, Y. Yu, "The capacity region for two classes of product broadcast channels," in *Proc. IEEE Int. Symp. Information Theory (ISIT2011),* pp.1544-1548, July 31-Aug. 5 2011.